\documentclass[paper]{revtex4}

\usepackage{graphicx}
\usepackage{axodraw}
\usepackage{epsfig}
\usepackage{fancyhdr}
\usepackage{empheq}
\usepackage{mathbbol}
\usepackage{wasysym}
\usepackage{pstricks}
\usepackage{color}
\usepackage{bbm}
\def\qq{\mathbbmss{q}}

\def\qu{{q_{_1}}}
\def\qd{{q_{_2}}}
\def\qub{{\bar q_{_1}}}
\def\qdb{{\bar q_{_2}}}

\newcommand{\xxx}[1]{{\color{red}\bf[#1]}} 

\begin{document}

\title{Higher Tetraquark Particles}

%

\author{NV Drenska$^{\dag,*}$, R Faccini$^{\dag, *}$} 
\author{AD Polosa$^*$}
\affiliation{$^\dag$Dipartimento di Fisica, Universit\`a di Roma `La Sapienza', Piazzale A Moro 2, Roma, I-00185, Italy\\
$^*$INFN Roma, Piazzale A Moro 2, Roma, I-00185, Italy}

\begin{abstract}
There are strong arguments favoring a four-quark interpretation of sub-GeV light scalar mesons and the 
diquark-antidiquark body-plan of the tetraquark seems to provide the most convincing picture. The building 
diquarks of these particles are assumed to be spin zero objects. In this paper we explore the possibility 
that radially excited aggregations of spin zero or spin one diquarks might exist and discuss the possibility
of the $Y(2175)$ state observed by BaBar and confirmed by BES being one such state.
\end{abstract}

\maketitle

\thispagestyle{fancy}




{\bf \emph{ Introduction}}. It has been shown recently how  the diquark-antidiquark interpretation
of the sub-GeV scalar meson nonet made of $f_0(980), a_0(980),\kappa(800),\sigma(500)$ can lead to a 
remarkable description of the decay properties of these  particles~\cite{hooft},
adding a rather strong confidence that they are indeed tetraquark objects.

In terms of diquarks $\qq$, the light scalar nonet can be interpreted as $\qq\bar\qq$ particles made of spin zero diquarks, 
so called ``good''. A good diquark operator in the attractive anti-triplet color channel (greek letters), 
antisymmetric in flavor (latin letters) can be written as~\cite{jaffe}:
\begin{equation}
\qq_{i\alpha}= \epsilon_{ijk}\epsilon_{\alpha \beta\gamma} \bar q^{j\beta}_c\gamma_5 q^{k\gamma}\label{defdq}
\end{equation}
Also ``bad'', spin one, diquarks can be conceived. A bad diquark operator can be written as:
\begin{equation}
\qq^{ij}_{\alpha}= \epsilon_{\alpha \beta\gamma} (\bar q^{\beta j}_c\vec\gamma  q^{k\gamma}+\bar q^{\beta k}_c\vec\gamma  q^{j\gamma}) 
\label{defdq2}
\end{equation}
Both represent positive parity, $0^+$ and $1^+$, states.  Similarly one can construct $0^-$ and $1^-$ operators as: $\bar q_c q$ and $\bar  q_c \vec \gamma \gamma_5q$. The latter are  identically zero in the `single mode configuration', quarks that are unexcited with respect to one another.
Lattice studies, see, e.g.~\cite{lucini}, suggest that diquarks are preferably (energetically) formed in  spin zero configurations.
In fact the most solid tetraquark candidates are scalars made of good diquarks.

The next step in building this new spectroscopy is to find states belonging to other multiplets. Recently,  the BaBar experiment 
has observed a new $J^{PC}=1^{--}$ resonance, the $Y(2175)$, decaying into $\phi f_0$~\cite{babar-y2175}, later confirmed by BES~\cite{bes-y2175}.
The fact that it has been first observed into a mode including a light scalar makes it a good candidate for  belonging to a higher tetraquark multiplet.

In this  paper we make a simple Ansatz that properly reproduces the light scalar nonet and utilize it to make predictions on possible excitations, focusing on $J^{PC}=1^{--}$ states and searching for a match with the $Y(2175)$.

{\bf \emph{ The Model}}. This is developed in the framework of a non-relativistic Hamiltonian including spin-spin interactions inside a single  diquark, spin-spin 
interactions between quarks and antiquarks belonging to the two diquarks forming the hadron, spin-orbit and a purely  orbital term:
\begin{equation}
H=2m_\qq + H_{_{SS}}^{(qq)} + H_{_{SS}}^{(q\bar q)}+H_{_{SL}}+H_{_{LL}}, 
\end{equation}
where:
\begin{eqnarray}
&&H_{_{SS}}^{(qq)}= 2\kappa_q (\vec S_\qu\cdot \vec S_\qd+ \vec S_{\qub}\cdot \vec S_{\qdb}) \notag\\ 
&&H_{_{SS}}^{(q\bar q)}=  2\kappa_{\qu\qdb}(\vec S_\qu\cdot \vec S_{\qdb}+\vec S_{\qub}\cdot \vec S_\qd)+2\kappa_{\qu\qub} \vec S_\qu\cdot \vec S_{\qub}+2\kappa_{\qd\qdb} \vec S_\qd\cdot \vec S_{\qdb}\notag\\
&&H_{_{SL}}= 2A_\qq(\vec S_\qq\cdot\vec L+\vec S_{\bar \qq}\cdot\vec L)\notag\\
&&H_{_{LL}}= B_\qq\frac{L(L+1)}{2}.
\label{ham}
\end{eqnarray}
The parameters in these equations are fit to data:  $m_\qq$ is the mass of the $[\qu\qd]$ diquark, $\kappa_q$ is the spin-spin coupling between the quarks inside the diquarks, 
$\kappa_{q_1q_2}$ are the spin-spin
couplings ranging outside the diquark shells,  $A_\qq$ is the diquark spin-orbit coupling, and  $B_\qq$ weights the
contribution of the total orbital angular momentum of the $\qq\bar\qq$ system to its mass; 
the overall factors of two are just conventional notations. We focus on the case where only light $u,d,s$ quarks are involved. 
The spin-spin interaction Hamiltonian has the form:
\begin{equation}
H_{_{SS}}=\sum_{\rm pairs} \frac{\kappa_{ij}}{m_i m_j}\,(\vec S_i \cdot \vec S_j)\, \delta^{3}(\vec r_{ij})
\label{eq:dgg}
\end{equation}
because the color-magnetic moments are inversely proportional to quark
masses. In Eqs.~(\ref{ham}) we incorporate the mass dependencies in the $\kappa_{ij}$ 
constants. 
The Hamiltonian~(\ref{eq:dgg}) describes  contact interactions. For this reason 
we could expect that allowing a relative orbital angular momentum  between the diquarks will decrease or switch-off the  spin-spin interactions between quarks and antiquarks, namely $H_{_{SS}}^{(q\bar q)}$. In the following we shall consider both cases.

The values of the couplings appearing in Eqs.~(\ref{ham}) were estimated in Ref.~\cite{x3872} from a fit to meson and baryon masses 
under the assumption that the spin-spin interactions are independent of whether the pair of quarks belong 
to a meson or a diquark. The estimates are summarized in Tab.~\ref{tab:par}.
\begin{table}[h]
\begin{center}
\begin{tabular}{|l||c|c|c|c|c|c|l||}
\hline
$[\qu\qd]$& $m_\qq $& $k_q$ & $k_{\qu\qdb}$ & $k_{\qu\qub}$ & $k_{\qd\qdb}$   \\ 
 \hline
$[qq]$&395&103&315&315&315 \\
\hline
$[sq]$&590&64&195&121&315 \\
\hline
$[ss]$&740&93&121&121&121 \\
\hline
\end{tabular}  \\ [1pt]
\caption{\footnotesize Estimate, in MeV, of the parameters in the Hamiltonian in Eq.~\ref{ham} 
depending on the diquark type $\qq=[\qu\qd]$. 
 \label{tab:par}}
\end{center}
\end{table}

Extending the same procedure to the $S=1,L=0,1$ meson states $\rho(770),a_1(1230),a_2(1320),b_1(1229)$~\cite{y4260} we also infer
the parameters related to the orbital angular momentum: $A_\qq=22.5$~MeV, $B_\qq=505$~MeV. 

To describe a $\qq \bar \qq$ quantum state we adopt the following non-relativistic notation:
\begin{equation}
|S_\qq,S_{\bar \qq};S_{\rm \qq\bar\qq}\rangle=|s^T\Gamma q, \bar s^T\Gamma \bar q; S_{\rm \qq\bar\qq} \rangle
\label{state}
\end{equation}
where $\Gamma\propto \sigma_2$ for a spin zero diquark and $\Gamma\propto \sigma_i$ for a spin one diquark. The action of a spin-spin interaction operator, e.g. $\vec S_{\bar s}\cdot \vec S_q$,  on
(\ref{state}) is described as follows:
\begin{equation}
(\vec S_{\bar s}\cdot \vec S_q) |s^T\Gamma q, \bar s^T\Gamma \bar q; S_{\qq\bar\qq}\rangle =
\frac{1}{4}\sum_j |s^T\Gamma \sigma_j q, \bar s^T \sigma_j^T\Gamma \bar q; S_{\qq\bar\qq}\rangle
\end{equation}

As an example let us diagonalize  the Hamiltonian in Eq.~(\ref{ham}) between 
scalars made of diquarks, i.e. $|0_\qq,0_{\bar \qq};0\rangle$ with a relative $L_{\qq\bar\qq}=0$. With an obvious 
shorthand notation:
\begin{equation}
|0_\qq,0_{\bar \qq};0\rangle = \frac{1}{2}|s^T\sigma_2 q,\bar s^T\sigma_2 \bar q;0\rangle := \frac{1}{2}\sigma_2\otimes \sigma_2.
\end{equation}
We can then compute:
\begin{equation}
\langle 0_\qq,0_{\bar \qq};0|\vec S_s\cdot \vec S_q |0_\qq,0_{\bar \qq};0\rangle = -\frac{1}{4}\times 3,
\end{equation}
where we have used the fact that $\sigma_j^T\sigma_2=-\sigma_2\sigma_j$ and $\sigma_j\sigma_j=3\times\mathbb{1}$. The final result is:
\begin{equation}
m=2m_{\qq}-3\kappa_q.
\end{equation}
If $\qq=[sq]$, then, using the values in Table~\ref{tab:par}, we get:
\begin{equation}
m=988~{\rm MeV}, 
\end{equation}
reproducing the mass of $m_{a_0}$ and $m_{f_0}$, considered as $[qs][\bar q\bar s]$ particles with the two diquarks in spin zero  and in $S-$wave~\cite{hooft}.
Repeating the same calculation with $\qq=[ud]$ one gets, for the $\sigma$-meson mass:
\begin{equation}
m=481~{\rm MeV}~.
\end{equation}

{\bf \emph{Higher Mass Tetraquark Spectrum}}. The next orbital excitation comes when $L_{\qq\bar\qq}=1$ and both good and bad diquarks are considered. Among these, also 
$1^{--}$ multiplets are generated, which are the main interest of this paper. 
To estimate the masses, one needs to repeat the diagonalization
with the basis:
\begin{eqnarray}
|1\rangle&=& |0_\qq,0_{\bar \qq};1_{J}\rangle\\
|2\rangle&=&\frac{|1_\qq,0_{\bar \qq};1_{J}\rangle+|0_\qq,1_{\bar \qq};1_{J}\rangle}{\sqrt{2}}\\
|3\rangle&=&|1_\qq,1_{\bar \qq};1_{J}\rangle
\end{eqnarray}
Since both the good and the bad  diquarks have positive parity, the state $|2\rangle$ has $P=C=-1$, provided that $L_{\qq\bar\qq}=1$.
For the states $|1\rangle$ and $|3\rangle$, since  
$C_{\qq\bar\qq}(-1)^{L_{\qq\bar\qq}}(-1)^{S_{\qq\bar\qq}}=1$, $C_{\qq\bar\qq}=-1$  provided that $S_{\qq\bar\qq}=0,2$ and $L_{\qq\bar\qq}=1$.

To perform the diagonalization we adopt the shorthand  notation described above:
\begin{eqnarray}
\label{eq:vector}
|0_\qq,0_{\bar \qq};1_{J}\rangle&=&\frac{1}{2}\sigma_2\otimes\sigma_2\notag\\
|1_\qq,0_{\bar \qq};1_{J}\rangle&=&\frac{1}{2}\sigma_2\sigma_i\otimes\sigma_2\notag\\
|0_\qq,1_{\bar \qq};1_{J}\rangle&=&\frac{1}{2}\sigma_2\otimes\sigma_2\sigma_i\notag\\
|1_\qq,1_{\bar \qq};1_{J}\rangle&=&\frac{1}{2\sqrt{2}} \epsilon^{ijk}\sigma_2\sigma_j\otimes\sigma_2\sigma_k,
\end{eqnarray}
Hence, it is rather straightforward to derive the mass term shift $\Delta m_{_{SS}}$ due to the part of the Hamiltonian in Eq.~(\ref{ham})
constraining only spin-spin interaction terms, $H_{_{SS}}$:
\begin{equation}
\Delta m_{_{SS}}=
\begin{bmatrix}
-3 \kappa_q & 0& 0\\
0&-\kappa_q-\kappa_{\qu \qdb}+(\kappa_{\qu\qub}+\kappa_{\qd \qdb})/2& 0\\
0&0&\kappa_q-\kappa_{\qu \qdb}-(\kappa_{\qu\qub}+\kappa_{\qd \qdb})/2
\end{bmatrix} 
\label{eq:matr}
\end{equation}
Writing the latter matrix as ${\rm \bf diag}(\lambda_1,\lambda_2,\lambda_3)$, 
the four  solutions for states having quantum numbers $1^{--}$ are: 
\begin{eqnarray}
&&m_Y^{(1)}(S_{\qq_1}=0,S_{\qq_2}=0,S_{\rm \qq\bar\qq}=0,L_{\qq\bar\qq}=1)=2 m_\qq +\lambda_1+ B_\qq\notag\\
&&m_Y^{(2)}(S_{\qq_1}=1,S_{\qq_2}=0,S_{\rm \qq\bar\qq}=1,L_{\qq\bar\qq}=1)=2 m_\qq +\delta+\lambda_2-2A_\qq+ B_\qq\notag\\
&&m_Y^{(3)}(S_{\qq_1}=1,S_{\qq_2}=1,S_{\rm \qq\bar\qq}=0,L_{\qq\bar\qq}=1)=2 m_\qq +2\delta+\lambda_3 -2A_\qq+ B_\qq\notag\\
&&m_Y^{(4)}(S_{\qq_1}=1,S_{\qq_2}=1,S_{\rm \qq\bar\qq}=2,L_{\qq\bar\qq}=1)=2 m_\qq +2\delta+\lambda_3-6A_\qq+ B_\qq
\label{eq:forms}
\end{eqnarray}
where $\delta=m_{\qq(S=1)}-m_{\qq(S=0)}$. Following Jaffe and Wilczeck~\cite{wilc}, 
we will assume for $\qq=[qq]$, $\delta\simeq 285$~MeV whereas for $\qq=[sq]$, $\delta\simeq 150$~MeV. 
The numerical values for the coefficients of $A_\qq$ and $B_\qq$, call them $a,b$,  are given in the folowing table. 
\begin{table}[h]
\begin{center}
\begin{tabular}{|l||c|c|c|l||}
\hline
&$ a(S_\qq,S_{\bar \qq},L) $& $b(S_\qq,S_{\bar \qq},L)$  \\ 
 \hline
$S_\qq=0,S_{\bar \qq}=0,L=1$&0&1 \\
\hline
$S_\qq=1,S_{\bar \qq}=0,L=1$&-2&1\\
\hline
$S_\qq=1,S_{\bar \qq}=1,S_{\qq\bar\qq}=2,L=1$&-6&1\\
\hline
$S_\qq=1,S_{\bar \qq}=1,S_{\qq\bar\qq}=1,L=1$&-2&1\\
\hline
$S_\qq=1,S_{\bar \qq}=1,S_{\qq\bar\qq}=0,L=1$&-2&1\\
\hline
\end{tabular}  \\ [1pt]
\caption{\footnotesize Eigenvalues of spin-orbit and angular momentum operators in~(\ref{ham}). 
All these combinations of diquark spins and orbital angular momenta allow a $J=L+S_{\qq\bar\qq}=1$ state.
 \label{tab:res}}
\end{center}
\end{table}

In case $\qq=[ss]$, only the last state in Eq.~(\ref{eq:vector}) is allowed since only bad diquarks can be formed by Fermi-Dirac. One should therefore consider only 
the $\langle 1_\qq,1_{\bar\qq};1_J|H_{_{SS}}|1_\qq,1_{\bar \qq};1_J\rangle$ correction to the mass, from Eq.~(\ref{eq:matr}) is equal to $\kappa_s-2\kappa_{s\bar s}$. We therefore have:
\begin{eqnarray}
&&m_Y^{(3)}(S_{\qq_1}=1,S_{\qq_2}=1,S_{\rm \qq\bar\qq}=0,L_{\qq\bar\qq}=1)=2m_\qq+2\delta+(\kappa_s-2\kappa_{s\bar s})-2A_\qq+B_\qq\notag\\
&&m_Y^{(4)}(S_{\qq_1}=1,S_{\qq_2}=1,S_{\rm \qq\bar\qq}=2,L_{\qq\bar\qq}=1)=2m_\qq+2\delta+(\kappa_s-2\kappa_{s\bar s})-6A_\qq+B_\qq.
\label{eq:forms2}
\end{eqnarray}
\begin{table}[h]
\begin{center}
\begin{tabular}{|l||c|c|c|c|c|l||}
\hline
               & $m_Y^{(1)} $& $m_Y^{(2)}$ & $m_Y^{(3)}$ & $m_Y^{(4)}$    \\
 \hline
$[qq^\prime]$&986&1432/1342&1293/1923&1203/1833 \\
\hline
$[sq]$&1493&1749/1726&1591/2004&1501/1914 \\
\hline
$[ss]$&--&--&2090/2333&2000/2243 \\
\hline
\end{tabular}  \\ [1pt]
\caption{\footnotesize Mass values $m_Y^{(i)}$ in MeV for the $1^{--}$ states  as computed from Eqs.~(\ref{eq:forms}) and (\ref{eq:forms2}).
When applicable, the first value includes spin-spin interactions between diquarks, the second one neglects them: $H_{_{SS}}^{(q\bar q)}=0$.
 \label{tab:miti}}
\end{center}
\end{table}
The numerical values for $m_Y^{(i)}$ masses can be found in Table~\ref{tab:miti}.
The fact that $1^{--}$ tetraquark particles require an angular momentum barrier $L_{\qq\bar\qq}\neq 0$ between diquarks must suppress the diquark-antidiquark chromomagnetic interactions.
Switching off spin-spin interactions between quarks and antiquarks  leads to the second estimates in Table~\ref{tab:miti}.

{\bf \emph{Tetraquark Decay Modes}}. From these results, if we want to identify the $Y(2175)$ as a tetraquark state, we have to resort to either a $\qq=[qs]$ hypothesis, with  spin-spin interactions between diquarks set to zero or to a $\qq=[ss]$ hypothesis: indeed we use the hypothesis that $f_0(980)$ in the decay products of $Y(2175)$ is itself a $\qq\bar \qq$ particle with $\qq=[qs]$. In order to test the match of the $Y(2175)$ with these assignments, we study its possible decays under both hypotheses, $\qq=[qs],[ss]$, and the 
correspondence with observations. 
With both assignments, the observed $Y(2175)\to\phi f_0(980)$  decay mechanism would be  
described by the following diagram:
\begin{center}
\SetScale{0.5}
\SetOffset(195,10)
\fcolorbox{white}{white}{
  \begin{picture}(285,154) (226,-119)
    \SetWidth{0.8}
      \Text(108,2)[l]{$s$}
     \Text(97,-14)[l]{$ s,q$}
      \Text(97,-28)[l]{$\bar s, \bar q$}
       \Text(108,-43)[l]{$\bar s$}
         \Text(45,-20)[l]{$Y(2175)$}
      \Text(260,-20)[l]{$f_0(980)$} 
      \Text(310,-20)[l]{$\phi$}    
        \Text(87,-8)[l]{$\qq$}
        \Text(87,-35)[l]{$\bar\qq$}
         \Text(221,-20)[l]{$q, s$}

    \ArrowLine(226,2)(346,2)
    \ArrowLine(346,2)(407,32)
    \ArrowLine(407,32)(600,32)
    \ArrowLine(227,-27)(349,-27)
    \ArrowLine(348,-27)(407,3)
    \ArrowLine(407,3)(497,3)
    \ArrowLine(347,-57)(226,-57)
    \ArrowLine(407,-87)(347,-57)
    \ArrowLine(496,-87)(407,-87)
    \ArrowLine(347,-87)(228,-87)
    \ArrowLine(404,-118)(347,-87)
    \ArrowLine(600,-118)(404,-118)
     \ArrowArcn(492.65,-40.71)(15.65,282.36,70.01)
    \Line(511,3)(511,-88)
     \Line(610,32)(610,-120)
  \end{picture}
}
\end{center}
\vspace{-2truecm}

As for other expected decays, a significant contribution is expected to come from $\phi\eta$ via the 
diagrams below:
\begin{center}
\SetScale{0.5}
\SetOffset(195,-20)
\fcolorbox{white}{white}{
  \begin{picture}(318,109) (193,-133)
    \SetWidth{0.8}
     \Text(90,-13)[l]{$s$}
     \Text(79,-27)[l]{$ s, q$}
      \Text(79,-51)[l]{$\bar s, \bar q$}
       \Text(90,-65)[l]{$\bar s$}
         \Text(35,-40)[l]{$Y(2175)$}  
        \Text(69,-19)[l]{$\qq$}
        \Text(69,-57)[l]{$\bar\qq$}
    \Text(260,-40)[l]{$\phi\eta(\phi\eta^\prime)$}  
    \ArrowLine(193,-27)(507,-27)
    \ArrowLine(195,-55)(318,-54)
    \Line(318,-54)(390,-102)
    \ArrowLine(390,-102)(509,-102)
    \ArrowLine(510,-56)(391,-56)
    \ArrowLine(316,-101)(198,-101)
    \ArrowLine(511,-131)(195,-131)
    \Line(391,-56)(316,-101)
  \end{picture}
}
\end{center}
\vspace{-1.3truecm}
Similar diagrams would also yield $\phi\phi$ and $\eta\eta$ decays, but they are forbidden by charge conjugation and Bose statistics 
selection rules.

We can also estimate the decay width of the  $Y(2175)\to\phi \eta$ channel. The decay proceeds through $P$-wave and
the matrix element is given by:
\begin{equation}
\langle \phi(p^\prime,\epsilon^{(\phi)})\eta(q)|Y(p,\epsilon^{(Y)})\rangle=g_V \epsilon^{\mu\nu\rho\sigma}p_\mu q_\nu \epsilon^{(\phi)}_\rho \epsilon^{(Y)}_\sigma
\end{equation}
The quark exchange amplitude considered above has been studied first in~\cite{scals} were a rather good fit
of the scalar meson decays to pseudoscalar, $S\to PP$, was obtained associating to this amplitude a coupling strength $A\simeq2.6$~GeV.  Discarding angular momentum barrier effects and following the definition given in~\cite{x3872}:
\begin{equation}
g_V M_V=\frac{A}{\sqrt{2}},
\end{equation}
where here $M_V=M_Y$, we get the following estimate for the $Y$ partial width in $\phi\eta$:
\begin{equation}
\Gamma(Y(2175)\to \phi\eta) = \xi\frac{A^2}{2}\frac{1}{8\pi M_Y^2} \frac{\sqrt{\lambda(M_Y^2,M_\phi^2,M_\eta^2)}}{2M_Y} \frac{(M_Y+M_\eta)^2-M_\phi^2}{M_Y^2}. 
\end{equation}
where $\xi=\frac{1}{6}$ or $\frac{2}{3}$ depending on the $\qq=[qs]$ or $\qq=[ss]$ respectively. We therefore estimate
$ \Gamma(Y(2175)\to \phi\eta) \sim 5 $ MeV or $\sim 20$ MeV under the two hypotheses respectively.

The most  typical decay mode expected for tetraquarks is the baryon-antibaryon one. 
Stretching the color string between the diquark and the antidiquark a $q\bar q$ pair is formed,  
$\qq\gluon q\bar q\gluon\bar\qq$, letting two baryons in the decay products. The favored decay diagram is
\begin{center}
\SetScale{0.5}
\SetOffset(195,10)
\fcolorbox{white}{white}{
  \begin{picture}(285,154) (226,-119)
    \SetWidth{0.8}
      \Text(108,2)[l]{$s$}
     \Text(97,-14)[l]{$ s,q$}
      \Text(97,-28)[l]{$\bar s, \bar q$}
       \Text(108,-43)[l]{$\bar s$}
         \Text(45,-20)[l]{$Y(2175)$}
      \Text(260,2)[l]{$\Xi, \Lambda$} 
      \Text(260,-42)[l]{$\bar \Xi, \bar \Lambda$}    
        \Text(87,-8)[l]{$\qq$}
        \Text(87,-35)[l]{$\bar\qq$}
         \Text(210,-21)[l]{$q, s$}
    \ArrowLine(226,2)(346,2)
    \ArrowLine(346,2)(407,32)
    \ArrowLine(407,32)(497,32)
    \ArrowLine(227,-27)(497,-27)%
    \ArrowLine(497,-57)(226,-57)
    \ArrowLine(347,-87)(228,-87)
    \ArrowLine(404,-118)(347,-87)
    \ArrowLine(497,-118)(404,-118)
     \ArrowLine(460,-85)(460,0)
      \ArrowLine(460,0)(497,0)
       \ArrowLine(497,-84)(460,-84)
    \Line(511,36)(511,-30)
     \Line(511,-54)(511,-124)
  \end{picture}
}
\end{center}
\vspace{-2truecm}
where the topology is suggested by the fact that the diquark in the $\Lambda$ baryon could only be of the $[ud]$ type.

Since the $\Xi\bar\Xi$ decay mode is phase-space forbidden for the $Y(2175)$, only the $\qq=[sq]$ assignation would allow a 
dominant baryonic decay, $Y(2175)\to\Lambda\bar\Lambda$.
With the other assignment, $\qq=[ss]$, the $Y(2175)\to\Lambda\bar\Lambda$ would be made possible  by the annihilation of an 
$s\bar{s}$ pair by, e.g., an instanton interaction giving two pairs of light quarks in the 
final state: in diagrammatic terms 
\begin{center}
\SetScale{0.5}
\SetOffset(195,-20)
\fcolorbox{white}{white}{
  \begin{picture}(319,109) (193,-133)
    \SetWidth{0.8}
         \Text(175,-30)[l]{${\cal I}$} 
    \ArrowLine(193,-27)(507,-27)
    \ArrowLine(195,-55)(318,-54)
    \Line(318,-54)(390,-102)
    \ArrowLine(390,-102)(509,-102)
    \ArrowLine(510,-56)(391,-56)
    \ArrowLine(316,-101)(198,-101)
    \ArrowLine(511,-131)(195,-131)
    \Line(393,-56)(318,-101)
    \SetWidth{0.5}
    \Vertex(356,-80){5.66}
    \SetWidth{0.8}
    \Line(358,-81)(392,-70)
    \ArrowLine(392,-70)(512,-71)
    \ArrowLine(510,-87)(392,-86)
    \Line(392,-86)(360,-80)
  \end{picture}
}
\end{center}
\vspace{-1.5truecm}
The 6-fermion instanton interaction has the form ${\cal L}_I\propto \det (\bar q_L^i q_R^j)$ and its role in scalar meson dynamics has been recently underscored in~\cite{hooft}. 

Nonetheless it is known from~\cite{hooft} that the instanton coupling, fitted to explain light scalar meson decays like $f_0(980)\to \pi\pi$, is about ten times smaller that the quark exchange one. 
Therefore, under the hypothesis that the baryonic mode would be instanton driven in the $\qq=[ss]$ case, we would not expect it to be easily visible. 

The latter diagram allows also a number of possible decays of the $Y(2175)$ like $p\bar{p}$, $\sigma\pi$, $\pi\pi\pi$, and 
$\eta\pi^0\pi^0$ in the $\qq=[sq]$ hypothesis and $K^-\kappa^+, K^- \pi^0 K^+$, and $\phi \pi^0\pi^0$ in the $\qq=[ss]$ one. 

\begin{figure}[htb]
\begin{center}
\epsfig{file=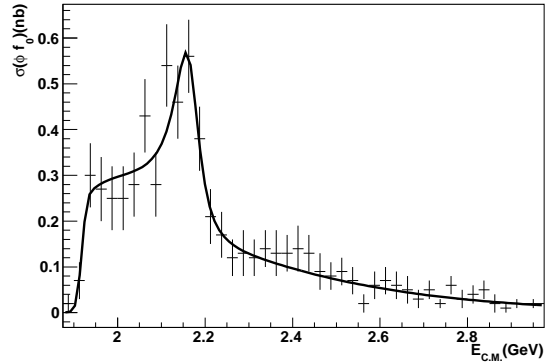,width=8cm}
\end{center}
\caption{Fit to the $\phi f_0$ invariant mass distribution.}
\label{fig:phif0}
\end{figure}

{\bf \emph{ Observed Y(2175) Decays}}.  In order to test the compatibility of the $Y(2175)$ state with a tetraquark interpretation and to discriminate between the 
two possible diquark compositions, we have reanalyzed the 
published BaBar data for $e^+e^-\to \phi f_0\gamma$~\cite{babar-y2175}, $\Lambda\bar{\Lambda}\gamma$~\cite{babar-ll},
and $\phi\eta\gamma$~\cite{babar-phieta}. These are initial state radiation processes, where $J^{CP}=1^{--}$ states
are produced together with the initial state photon. The invariant mass of the system produced with the 
photon is then  expected to show a resonant behavior in correspondence to states.

We perform simultaneous fits applying a consistent notation for the Breit-Wigner and 
several possible models for the non-resonant component $A_{nr}$. The general notation for the 
expected cross section as a function of the invariant mass of the system under study is 
\begin{eqnarray}
\sigma(m)&\propto& \Phi^f_{PS}(m)|A(m)|^2\\ \notag
A(m)&=&e^{i\delta}A_{nr}(m)+\sqrt{\sigma_0 {\cal{B}}(Y\to f)} \frac{m_0\Gamma_{\rm tot}(m_0)}{m^2-m_0^2+i\Gamma_{\rm tot}(m)m_0}
\end{eqnarray}
where $\delta$ is the relative phase between the two components at the pole; $\Phi^f_{PS}(m)=(p(m)/p(m_0))^{\alpha_f}$ is the final state dependent phase space factor: $\alpha_{\phi\eta}=3$, $\alpha_{\phi f_0}=\alpha_{\Lambda\bar{\Lambda}}=1$; $p(m)$ is the momentum of the two particles in the final state when their c.o.m. energy is $m$;
$m_0$ and $\sigma_0$ are the pole mass and production cross section and are independent of the considered 
final state; ${\cal B}(Y\to f)$ is the Branching Fraction to the specific final state; $\Gamma_{\rm tot}(m)$ is 
the comoving width,  the sum over the considered final states plus a constant term to account 
for all other decays with thresholds far from the pole mass:
\begin{eqnarray}
\Gamma_{\rm tot}(m)&=& \Gamma_0\left(1-\sum_f {\cal{B}}(Y\to f) (1-\xi^f(m) )\right)\\ \notag
\xi^f(m) &=& \Phi^f_{PS}(m)/  F_{BW}(m) 
\end{eqnarray}
where $F_{BW}$ is the Blatt-Weisskopf factor~\cite{BW} and $\Gamma_0$ is the bare width. 
Note that for masses below the threshold of a given final state the corresponding $\xi$ is imaginary.

\begin{table}[bht]
\begin{center}
\caption{ Fit results to the $\phi f_0$ invariant mass distribution (Fit A), the $\phi f_0$ and $\Lambda \Lambda$
invariant mass distributions (Fit B), and the invariant mass distributions of all three modes (Fit C). The meaning of the symbols is 
explained in the text.
\label{tab:fits}}
\begin{tabular}{|c||c|c|c|c|c|}
\hline
Fit & $\chi^2/$DOF&$m_0$ (MeV) & $\Gamma_0$ (MeV) & $R(\Lambda\bar{\Lambda},\phi f_0)$  & $R(\phi\eta,\phi f_0)$ \\ \hline
A  & 25/36&$2167\pm11$ & $69\pm21$ & N/A & N/A \\ 
B  & 47/45&$2158\pm11$ & $66\pm20$ &  $44\pm19$  & N/A \\ 
C  & 85/103&$2153\pm 9$ & $72\pm 20$& $6.6\pm3.5$ & $10\pm3$ 
\\ \hline
\hline
\end{tabular}
\end{center}
\end{table}
Of the three considered modes, in two cases the $Y(2175)$ would decay to states where 
it is above threshold while for the other ($\Lambda\bar{\Lambda}$) $m_0$ is below threshold.    
A fit to the discovery mode, $\phi f_0$, with 
$A_{nr}=A\times e^{-k(E-m_{\phi}-m_{f_0})}\times (1-e^{-(E-m_{\phi}-m_{f_0})^4/a_1})$ (see Fig.~\ref{fig:phif0})
 returns the results listed as ``Fit A'' in Tab.~\ref{tab:fits}.

\begin{figure}[htb]
\begin{center}
\epsfig{file=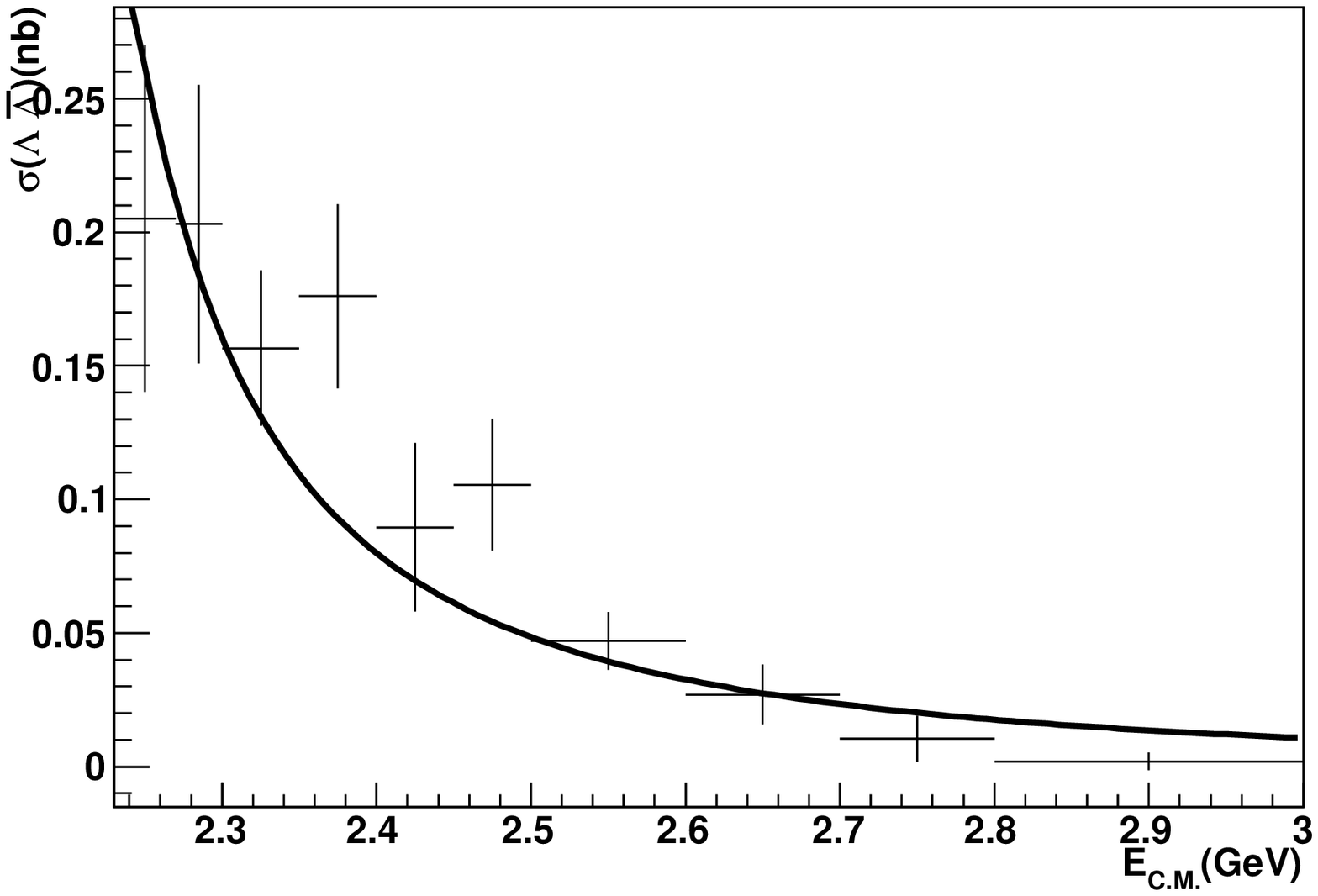,width=7cm}
\epsfig{file=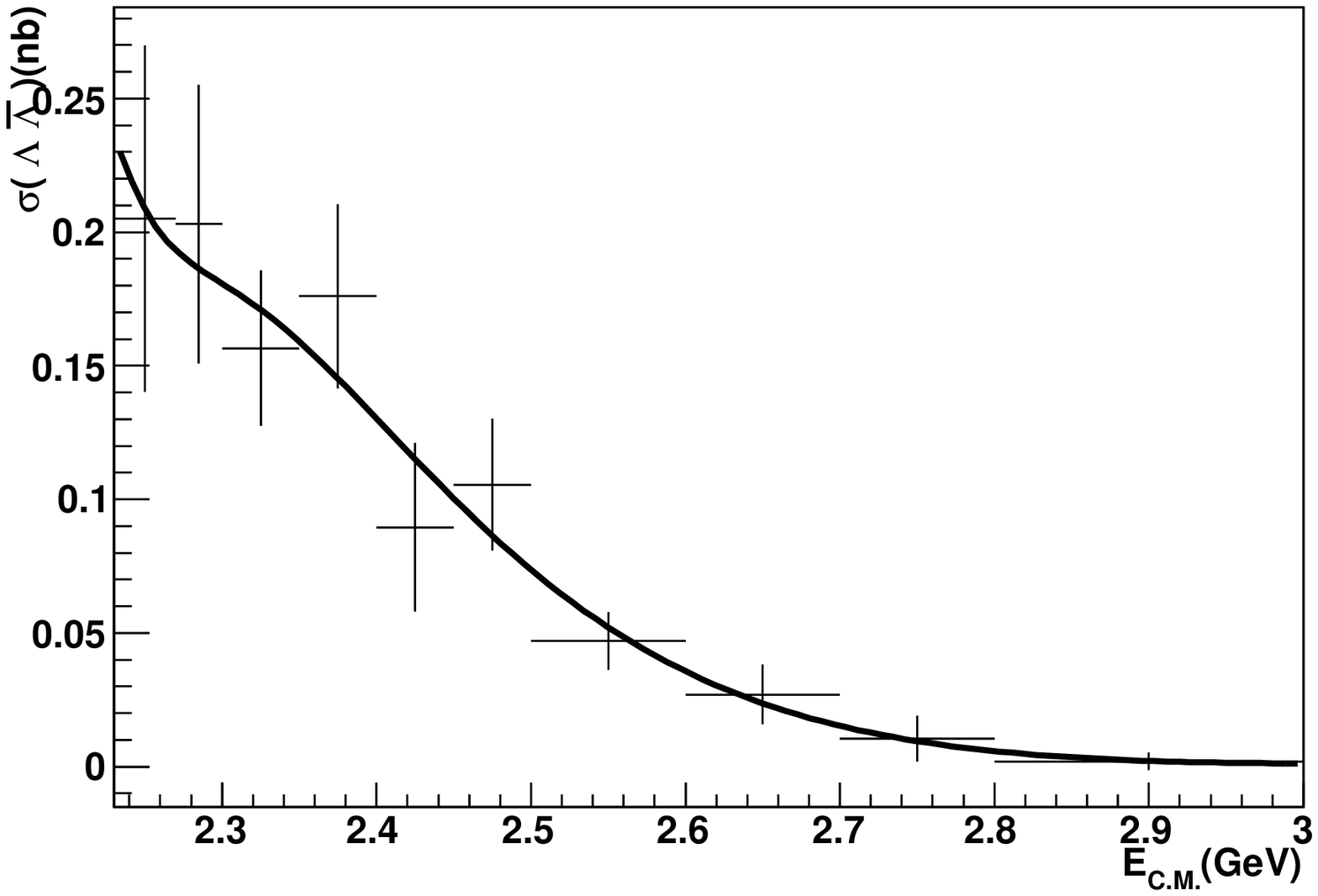,width=7cm}
\end{center}
\caption{Fit to the $\Lambda\bar{\Lambda}$ invariant mass distribution under the assumption of no non-resonant 
contribution (left), or assuming an exponential amplitude for it (right).}
\label{fig:ll}
\end{figure}

Including the $\Lambda \bar \Lambda$ mode in a simultaneous fit, with common bare mass and width
and letting the ratio between the branching fractions 
$R(\Lambda\bar{\Lambda},\phi f_0)={\cal{B}}_{\Lambda\bar{\Lambda}}/{\cal{B}}_{\phi f_0}$ float, 
 yields different results depending on the assumptions  $A_{nr}$ in the 
$\Lambda\bar{\Lambda}$ mode (see Fig.~\ref{fig:ll}). In case we assume no non-resonant contribution,
$A_{nr}=0$, we find that the $Y(2175)\to\Lambda\bar{\Lambda}$ decay can explain 
the whole observed spectrum,
with mass and width parameters consistent with the fit to the $\phi f_0$ mode. The results are reported as 
``Fit B'' in Tab.~\ref{tab:fits}

There is an indication that the $\Lambda\bar{\Lambda}$ decay is favored, even by one order of magnitude.
If instead we assume $A_{nr}(m)=A\times e^{-km}$, letting $k$ float,
we observe no significant decay into $\Lambda\bar{\Lambda}$, but with huge uncertainties:
$R(\Lambda\bar{\Lambda},\phi f_0)=23\pm51$.

\begin{figure}[htb]
\begin{center}
\epsfig{file=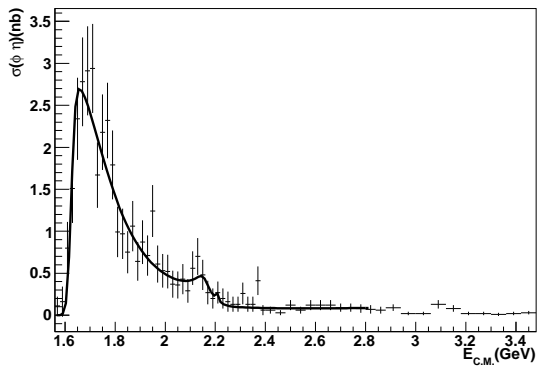,width=8cm}
\end{center}
\caption{ $\phi f_0$, $\Lambda\bar\Lambda$, and $\phi \eta$ invariant mass distributions with the simultaneous fit overlaid.}
\label{fig:phieta}
\end{figure}

Finally, the $\phi\eta$ mass distribution shows a significantly higher background than the other modes
 and small structure in a position which is lower and narrower than the one observed in the $\phi f_0$
channel and marginally consistent with it. 
Fitting the three distributions simultaneously as shown in Tab.~\ref{tab:fits}, ``Fit C'' gives an
 overall good fit (see Fig.~\ref{fig:phieta}).

We can then conclude that the presence of the 
$Y(2175)\to\Lambda\bar{\Lambda}$ is is suggested by the 
fact that its existence would explain the whole $\Lambda\bar{\Lambda}$ mass spectrum. Under this hypothesis the baryonic decay mode 
would be dominant (${\cal B}_{\Lambda\bar\Lambda}/{\cal B}_{\phi f_0}=44\pm19$), thus favouring $\qq=[qs]$ for the $Y(2175)$.
 As for the $\phi\eta$ mass distribution, uncertainties are large, but the case of a relatively large decay amplitude into it is not 
disfavored.

Both the calculation of the mass spectrum and the reanalysis of the experimental data tend to favor the assignment of the $Y(2175)$ to
a $[sq][\bar s\bar q]$ state with both diquarks in the $S=1$ state and with one unit of relative orbital angular momentum.   

{\bf \emph{Conclusions}}. In this paper we have studied the consequences of allowing spin one diquarks to build $1^{--}$ $(\qq\bar\qq)$ orbitally excited tetraquark states potentially visible in processes with initial state radiation at BaBar and Belle. In particular we have focused on the $Y(2175)$ resonance recently discovered by BaBar. This particle could be the first tetraquark state showing the expected baryon-antibaryon decay.
Indeed, reanalyzing BaBar data, we find that if we set to zero the non-resonant contributions, the $\Lambda\bar \Lambda $ decay mode is the prominent one, indicating a $\qq=[qs]$ assignation for the $Y(2175)$.

\begin{acknowledgments}
We wish to thank Gino Isidori for his comments and suggestions on the manuscript and Luciano Maiani for many discussions and fruitful collaboration.
\end{acknowledgments}

\bigskip 

\end{document}